\documentstyle[aps,epsf,rotate,citesort,preprint]{revtex}

\begin{document}

\draft
\title{Multifractal Properties of Price Fluctuations\\ of Stocks and
Commodities}

\author{Kaushik~Matia$^1$, Yosef~Ashkenazy$^2$ and
H. Eugene~Stanley$^1$}

\address{$^1$Center for Polymer Studies and Department of Physics\\
Boston University, Boston, MA 02215\\ $^2$Department of Earth, Atmosphere, and
Planetary Sciences\\ MIT, Cambridge, MA 02139}

\maketitle

\begin{abstract}

We analyze daily prices of 29 commodities and 2449 stocks, each over a
period of $\approx 15$ years.  We find that the price fluctuations for
commodities have a significantly broader multifractal spectrum than for
stocks. We also propose that multifractal properties of both stocks and
commodities can be attributed mainly to the broad probability
distribution of price fluctuations and secondarily to their temporal
organization. Furthermore, we propose that, for commodities, stronger
higher order correlations in price fluctuations result in broader
multifractal spectra.

\end{abstract}




The study of economic markets has recently become an area of active
research for physicists \cite{Books}, in part because of the large
amount of data that can be accessed for statistical analysis. Markets
are complex systems for which the variables characterizing the state of
the system---e.g., the price of the goods, the number of trades, and
the number of agents, are easily quantified. These variables serve as
fundamental examples of scale invariant behavior---the scaling laws are
valid for time scales from seconds to decades.

Much interest has concentrated on stocks, where a number of empirical
findings have been established, such as \cite{Works} (i) the
distribution of price changes is approximately symmetric and decays with
power law tails with an exponent $\alpha + 1 \approx 4$ for the
probability density function; (ii) the price changes are exponentially
(short-range) correlated while the absolute values of price changes
(``volatility'') are power-law (long-range) correlated.

Unlike stock and foreign exchange markets, commodity markets have
received little attention. Recently, it was found \cite{Kaushik} that
commodity markets have qualitative features similar to those of the
stock market. This similarity is intriguing because the commodity market
has special features such as: (i) most commodities require storage; (ii)
most commodities require transportation to bring them to the market from
where they are produced; and (iii) it is plausible that commodities may
exhibit a slower response to change in demand because the price depends on
the supply of the actual object.

The multifractal (MF) spectrum \cite{Gene} reflects the $n$-point
correlations~\cite{Barabasi} and thus provides more information about
the temporal organization of price fluctuations than 2-point
correlations. Previous work reports a broad MF spectrum of stock indices
and foreign exchange
markets~\cite{Mandelbrot,Bershadskii,Strutzik,Fisher,Sun,Canessa,Ausloos}. Two
recent models \cite{Bacry,Pochart} explain the observed MF properties by
assuming that price changes are the product of two stochastic variables,
one being uncorrelated and normally distributed and the other being
correlated and log-normally distributed. The price changes predicted by
these models do not have the power law probability distribution
\cite{Works,Kaushik} observed empirically, and thus shuffling the price
changes significantly narrows the MF spectrum.

Here we show that the MF properties of commodities and stocks result
partly from the temporal organization and partly from the power law
distribution of price fluctuations. We also conjecture that it is
feature (iii) of commodity markets that leads to a broader MF spectrum
of commodities compared to stocks.

We define the normalized price fluctuation (``return'') as
\begin{equation}
g(t) \equiv \frac{\ln S(t+\Delta t)- \ln S(t)}{\sigma},
\label{e.return}
\end{equation}
where here $\Delta t = 1$ day, $S(t)$ is the price, and $\sigma$ is the
standard deviation of $\ln S(t+\Delta t)- \ln S(t)$ over the duration of
the time series (typically 15 years). We use the multifractal detrended
fluctuation analysis (MF-DFA) method~\cite{Jan} to study the MF
properties of the returns for stocks and commodities.
Following~\cite{Jan}, given a time series $x_k$ we execute the following
steps: (i) Calculate the profile; $Y_i \equiv \sum_{k=1}^{i}
[x_k-\langle x \rangle ]~~~i=1,...,N$, where $N$ is the length of the
time series, and $\langle x \rangle$ is the mean. (ii) Divide $Y_i$ into
$N_s \equiv int(N/s)$ segments. (iii) Calculate the local trend
$y_{\nu}(i)$ for segments $\nu=1,...,N_s$ by least square polynomial
fit. (iv) Determine the mean square fluctuation in each segment
$F_2(s,\nu) \equiv \frac{1}{s} \sum_{i=1}^{s} (|Y_{(\nu-1)s+i} -
y_{\nu}(i)|)^2$. (v) Evaluate $F_q(s) \equiv \frac{1}{N_s}
\sum_{\nu=1}^{N_s} F_2(s,\nu)^{q/2}$. The scaling function of moment
$q$, $F_q(s)$ \cite{Jan} follows scaling law
\begin{equation}
F_q(s)\sim s^{\tau(q)}.
\label{part-fn2}
\end{equation}
Negative $q$ values weight more small fluctuations, while positive
values of $q$ give more weight to large fluctuations.

When the contribution of the small fluctuations is comparable to the
contribution of the large fluctuations, the series is {\it monofractal}
and $\tau(q)=h q$ where $h=const$ is the Hurst exponent. If $\tau(q)$
nonlinearly depends on $q$, the series is MF. The Legendre transform of
$\tau(q)$ is
\begin{equation}
f(h)\equiv q h - \tau(q)
\qquad\qquad\mbox{where}\qquad\qquad h\equiv\frac{d\tau(q)}{d q}.
\label{legender}
\end{equation}
Monofractal series have only one value of $h$, unlike MF series
which have a distribution of $h$ values.

\begin{figure}
\narrowtext
\vspace*{0.05cm}
\vspace*{-0.9cm}
\centerline{
\epsfysize=1.0\columnwidth{\rotate[r]{{\epsfbox{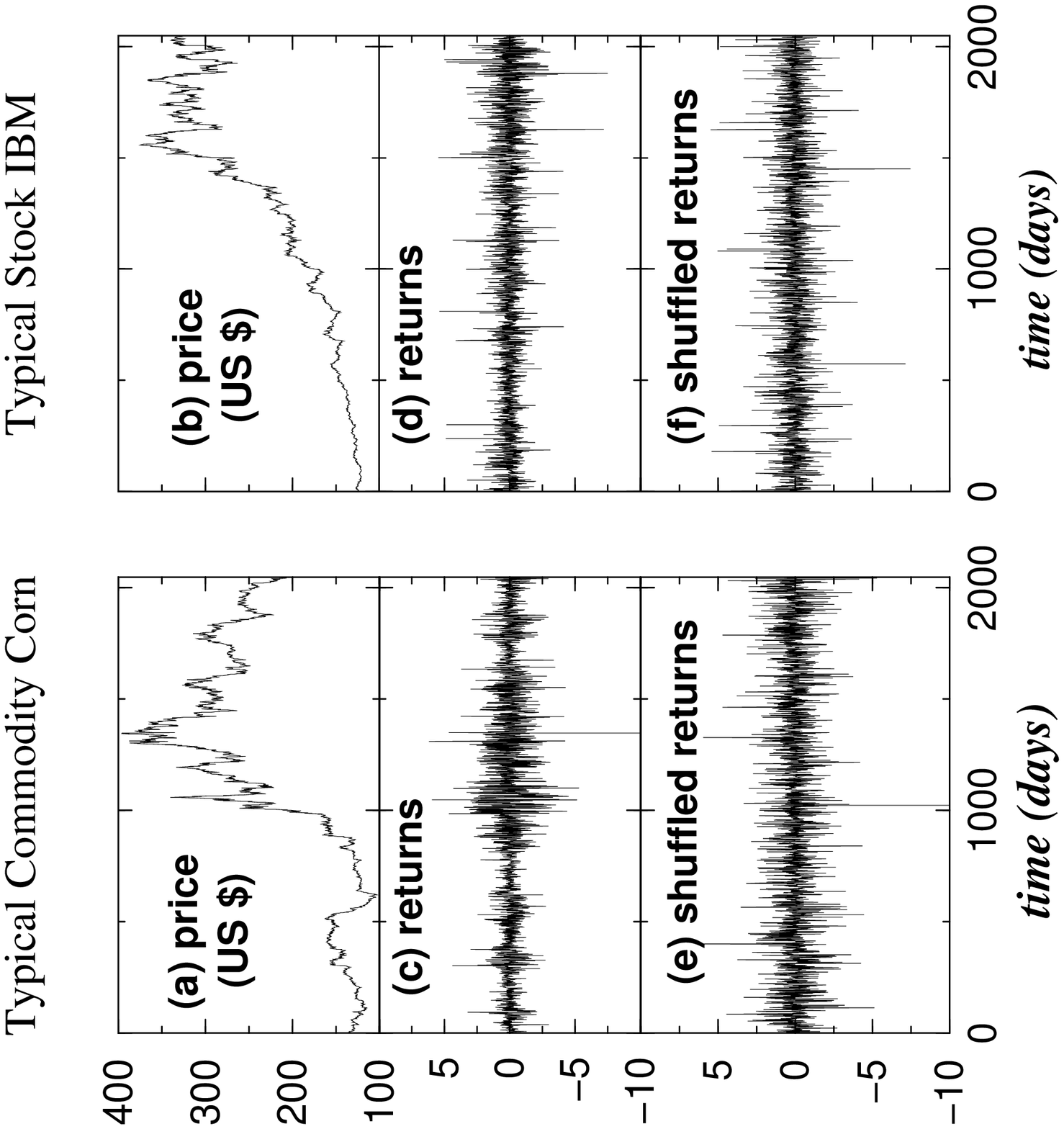}}}}
}
\vspace*{0.6cm}
\caption{Analysis of 2048 daily returns covering the time period May
1993 -- June 2001. Price of (a) corn, a typical commodity, and (b) IBM,
a typical stock. Returns for (c) corn and (d) IBM; unlike stocks,
commodity returns appear more clustered into ``patches'' of small and
large fluctuations, even though there are no 2-point correlations
(Fig.~2). Returns for (e) corn and (f) IBM after shuffling the
returns; because all $n$-point correlations are now removed, stocks and
commodities look similar and do not appear to cluster into ``patches.''}
\label{Fig1}
\end{figure}

We analyze a database consisting of daily prices of 29
commodities~\cite{Cdatabase} and 2449 stocks~\cite{Sdatabase} spread
over time periods ranging from 10 to 30 years (the average period
is $\approx$ 15 years). Figure~\ref{Fig1} shows the price of a typical
commodity, corn, and a typical stock, IBM, and their corresponding
returns. One striking difference between the commodity and the stock is
that the commodity returns are more ``clustered'' into patches of small
and large fluctuations. This feature is not reflected in the
distribution and the autocorrelation function of the price
fluctuations~\cite{Kaushik}.

To emphasize the clustering of commodities, we shuffle the returns by
randomly exchanging pairs, a procedure that preserves the
distribution of the returns but destroys any temporal correlations.
Specifically, the shuffling procedure consists of the following steps

\begin{itemize}

\item[{(i)}] Generate pairs $(m,n)$ of random integer numbers (with $m,n
\leq N$) where $N$ is the total length of the time series to be
shuffled.

\item[{(ii)}] Interchange entries $m$ and $n$.

\item[{(iii)}] Repeat steps (i) and (ii) for $20 N$ steps. (This step
ensures that ordering of entries in the time series is fully shuffled.)

\end{itemize}

\noindent
To avoid systematic errors caused by the random number generators, the
shuffling procedure is repeated with different random number seeds for
each of the 2449 stocks and 29 commodities. The shuffled commodity
series of necessity loses its clustering \cite{Shuffling-note}
[Fig.~\ref{Fig1}(e)]. On the other hand, the shuffled stock series
resembles the original one.

\begin{figure}
\narrowtext
\vspace*{-0.9cm}
\centerline{
\epsfysize=0.90\columnwidth{\rotate[r]{{\epsfbox{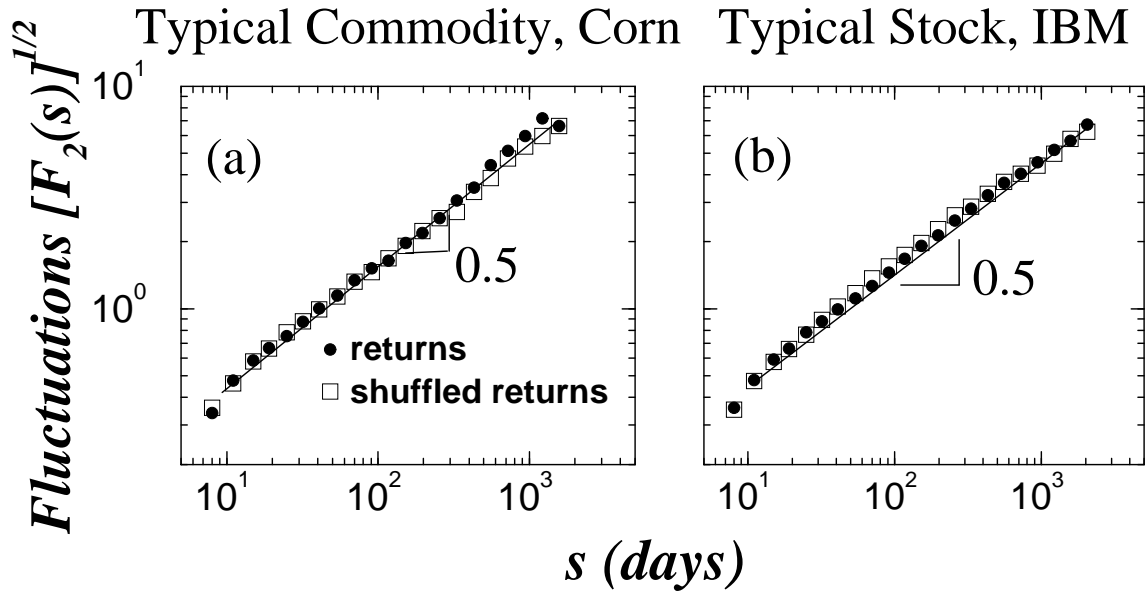}}}}
}
\vspace*{0.6cm}
\caption{DFA analysis of two-point correlations of returns of (a) a
typical commodity (corn) and (b) a typical stock (IBM), before and after
shuffling. The exponent of 0.5 indicates that both the commodity and the
stock are uncorrelated in time, so the two-point correlation function
does not provide information regarding the clustering into patches
(Fig.~1).}
\label{Fig2}
\end{figure}

First we compare the 2-point correlations using DFA~\cite{Peng} of the
shuffled and the unshuffled returns for commodities and stocks
(Fig.~\ref{Fig2}). Both corn (and all other commodities analyzed) and
IBM (and all other stocks analyzed) have returns uncorrelated for time
scales larger than one day~\cite{Works,Kaushik}. Thus, studying the
2-point correlations is not sufficient to uncover the clustering of the
commodity returns.

\begin{figure}
\narrowtext
\vspace*{-0.9cm}
\centerline{
\epsfysize=0.65\columnwidth{\rotate[r]{{\epsfbox{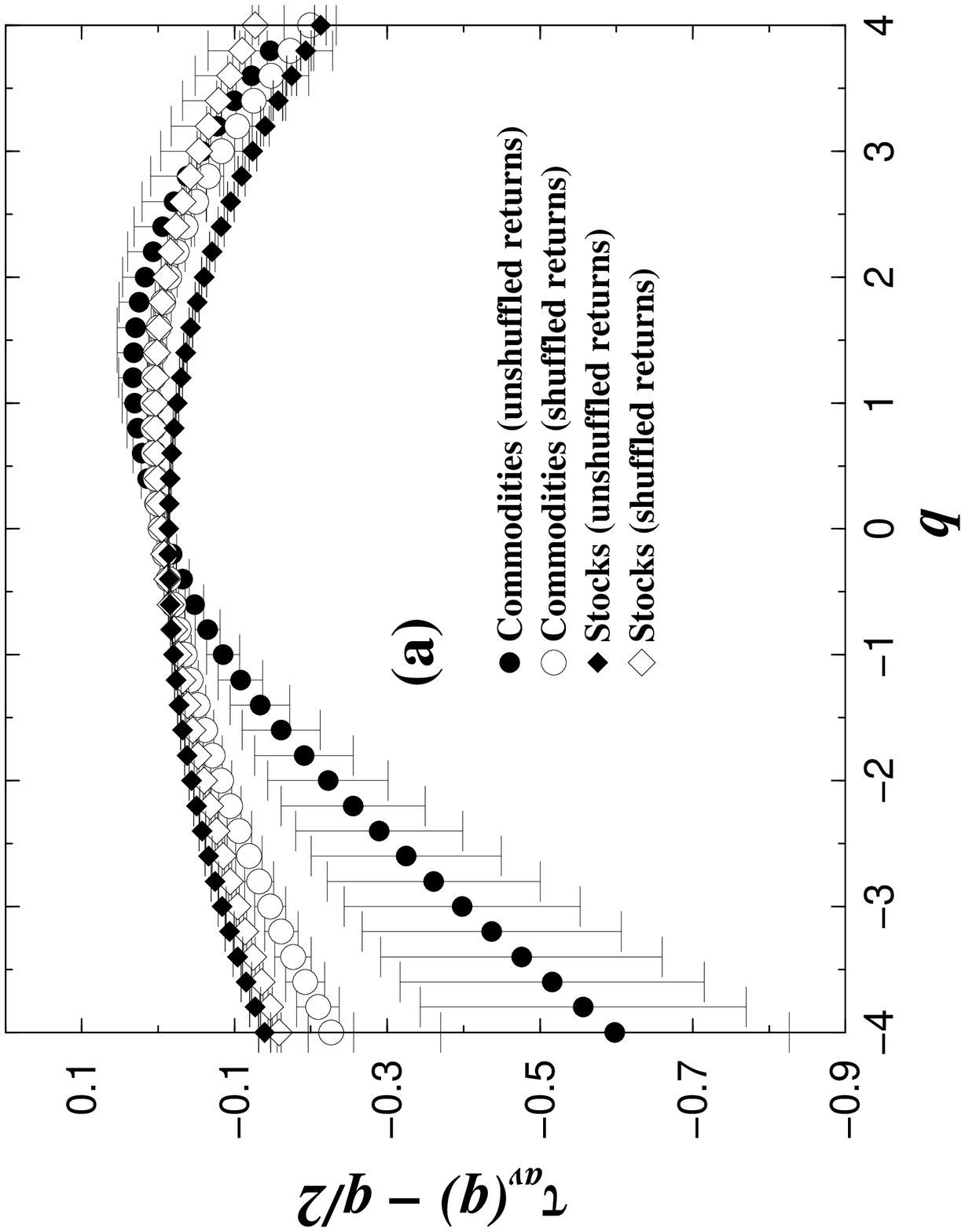}}}}
}
\centerline{
\epsfysize=0.65\columnwidth{\rotate[r]{{\epsfbox{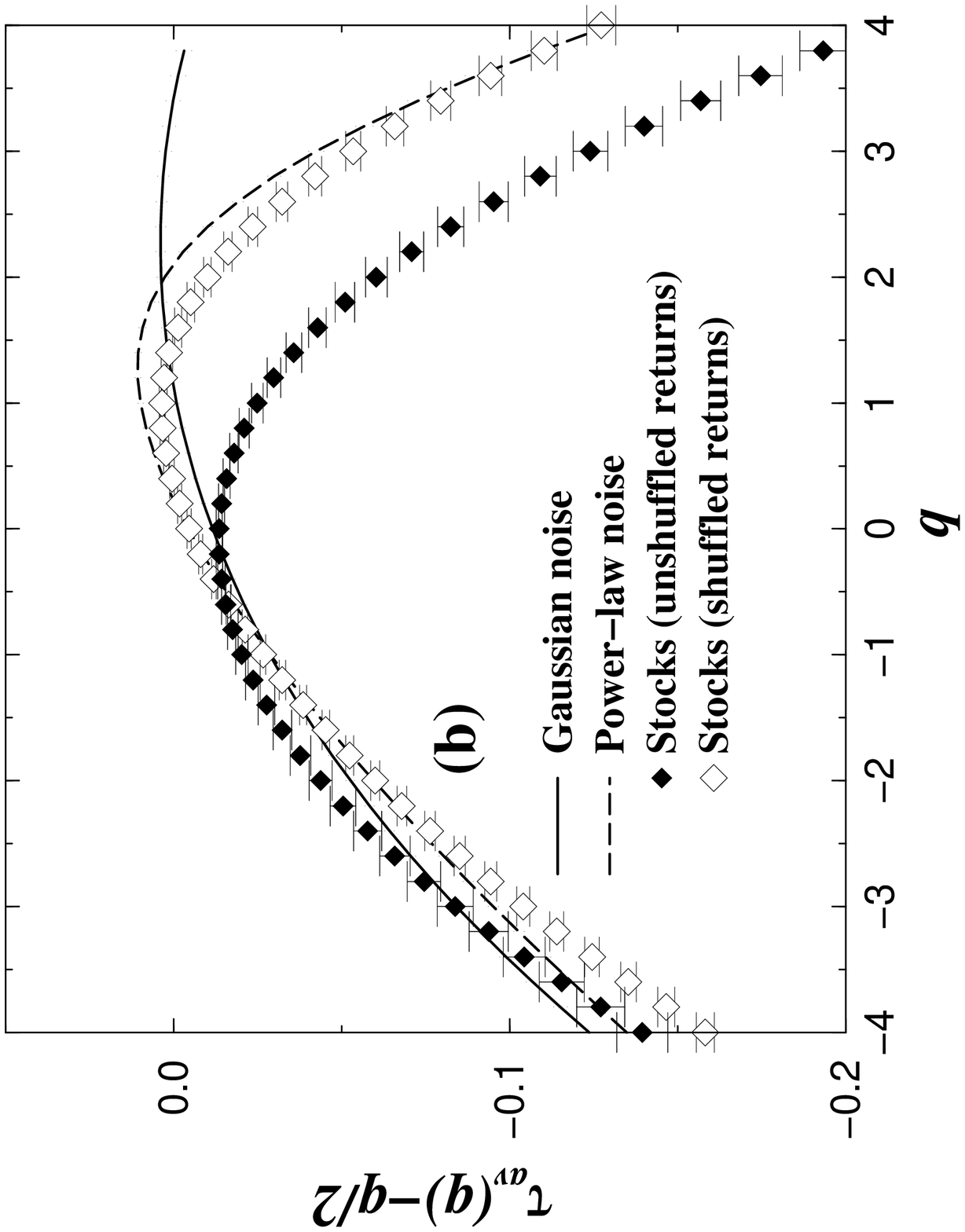}}}}
}
\vspace*{0.6cm}
\caption{(a) $\tau_{\mbox{\scriptsize av}}(q)$ for returns and shuffled
returns for 29 commodities and 2449 stocks. To better visualize the
results we plot $\tau_{\mbox{\scriptsize av}}(q)-q/2$ instead of
$\tau_{\mbox{\scriptsize av}}(q)$. The exponents
$\tau_{\mbox{\scriptsize av}}(q)$ are calculated for window scales of
$10-100$ days. After shuffling $\tau_{\mbox{\scriptsize av}}(q)$ are
comparable for both stocks and commodities. (b) $\tau_{\mbox{\scriptsize
av}}(q)$ spectrum of the returns and shuffled returns for stocks,
compared with uncorrelated surrogate data with Gaussian probability
distributions and power law probability distributions (with power law
exponent $\alpha \approx 3$). After shuffling, $\tau_{\mbox{\scriptsize
av}}(q)$ for stocks becomes comparable with $\tau_{\mbox{\scriptsize
av}}(q)$ of the surrogate data obtained for the power law probability
distribution.}
\label{Fig3}
\end{figure}

Next, we analyze the MF properties of the returns of stocks and
commodities. Figure~\ref{Fig3}(a) displays separately the averages
\begin{equation}
\tau_{\mbox{\scriptsize
av}}(q)\equiv\frac{1}{N}\sum_{i=1}^{N}\tau_i(q),
\end{equation}
for $N=29$ commodities, and for $N=2449$ stocks. Note that (i) the
scaling exponents, $\tau_{\mbox{\scriptsize av}}(q)|_{q<0}$ for
commodities and stocks significantly differ, whereas
$\tau_{\mbox{\scriptsize av}}(q)|_{q>0}$ are similar, suggesting that
commodities are similar to stocks for the large fluctuations and they
differ for the small fluctuations~\cite{Note2}; (ii) we observe that
after shuffling the returns, $\tau_{\mbox{\scriptsize av}}(q)$ for
stocks hardly changes for $q<0$, but $\tau_{\mbox{\scriptsize av}}(q)$
for commodities changes and becomes comparable to stocks for the entire
range of $q$~\cite{Pfn-note}.

In order to study the contribution of the power law tails of the returns
on the MF spectrum we generate surrogate data sets (i) with a normal
distribution and (ii) with power law tails with $\alpha\approx 3$ (as is
observed empirically~\cite{Works,Kaushik}). Figure~\ref{Fig3}(b)
displays $\tau_{\mbox{\scriptsize av}}(q)$ averaged over 2449
realizations of surrogate data, each with 3000 data points.  The
$\tau_{\mbox{\scriptsize av}}(q)$ of the surrogate power law distributed
data is very close to the $\tau_{\mbox{\scriptsize av}}(q)$ of stocks
after shuffling. This indicates that a significant part of the
$\tau_{\mbox{\scriptsize av}}(q)$ spectrum of stocks and commodities
comes from the power law distribution of the returns. Note that there is
a small difference in $\tau_{\mbox{\scriptsize av}}(q)$ of stocks before
and after shuffling, indicating that the power-law distribution of the
returns is not the only source of multifractality but that there is also
a relatively small contribution due to temporal organization of
returns. For commodities this temporal organization is more dominant.

\begin{figure}
\narrowtext
\vspace*{-0.9cm}
\centerline{
\epsfysize=0.65\columnwidth{\rotate[r]{{\epsfbox{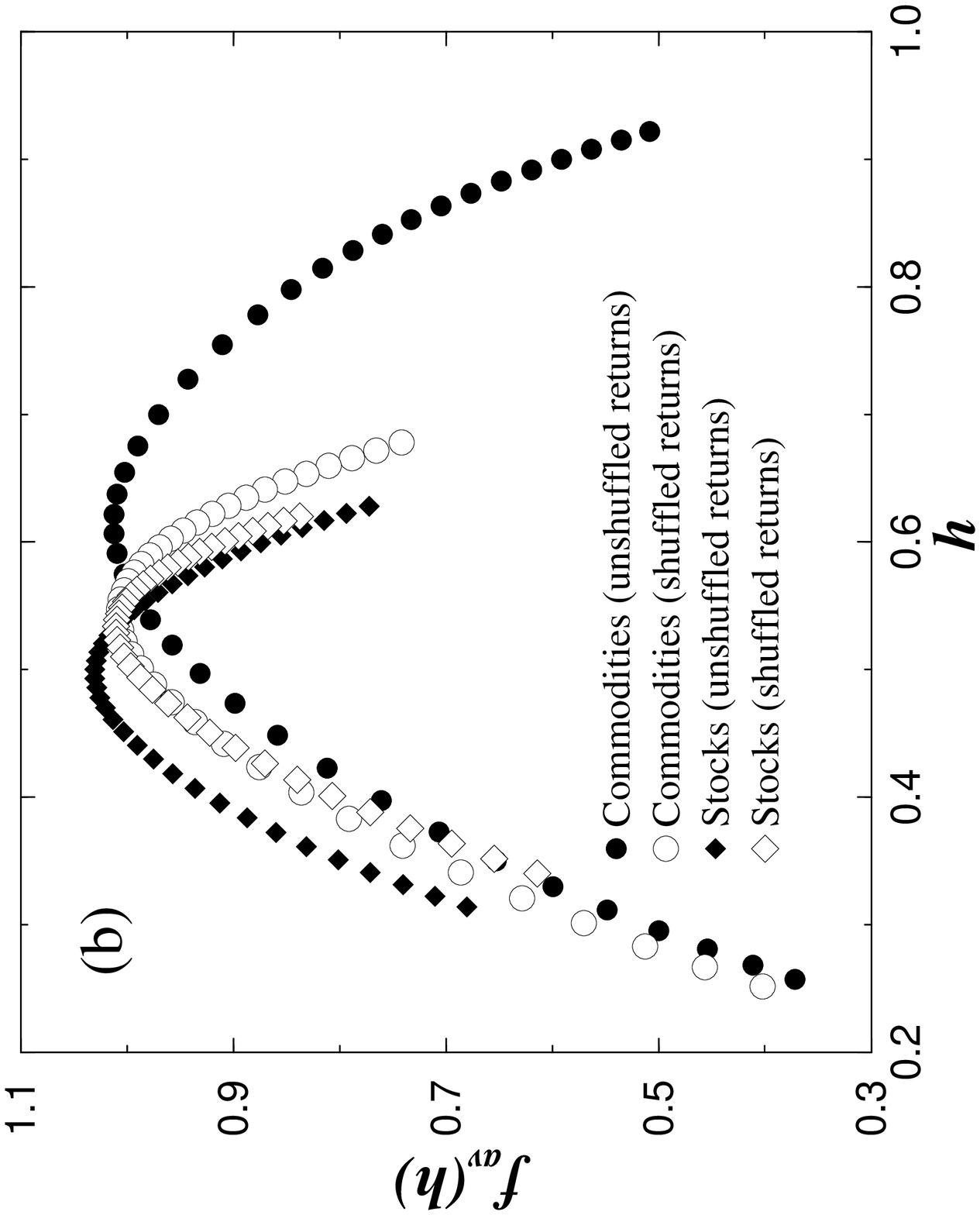}}}}
}
\vspace*{0.6cm}
\caption{$f_{\mbox{\scriptsize av}}(h)~vs.~h$ computed from the
$\tau_{\mbox{\scriptsize av}}(q)~vs.~q $ for commodities and
stocks. Shown are also both unshuffled and shuffled returns. For
commodities, the unshuffled returns show a broader MF spectrum than for
the shuffled returns, consistent with the hypothesis that the broad MF
spectrum arises due to temporal organization and the broad power law
distribution of price fluctuations (cf. Fig.~3b).}
\label{Fig4}
\end{figure}

Figure~\ref{Fig4} displays the MF spectrum of the unshuffled and
shuffled returns for commodities and stocks. The temporal organization
of commodity returns is reflected in the fact that the MF spectrum for
unshuffled commodities is broader than for shuffled commodities and
stocks.


Demand fluctuations drive price fluctuations, and it is plausible that
stocks respond more quickly than commodities to demand changes.
Stochastic perturbations, together with the immediate price response to
demand changes, may weaken the existing higher order temporal
organization, which may be the reason for less clustering for stocks
(Fig.~\ref{Fig1}). Commodities, on the other hand, have a slower
response. Thus, small or short-time perturbations are ``felt'' less by
commodities than by stocks. This scenario is consistent with the
appearance of patches of small commodity fluctuations followed by
patches of large commodity fluctuations (Fig.~\ref{Fig1}). We conjecture
that the more homogeneous returns of stocks explain the difference
between the MF properties of stocks and commodities.

In summary, we find that commodities have a broader MF spectrum than
stocks. A major contribution to multifractality is the power law tail of
the probability distribution of the returns.  Moreover, the MF spectra
of stocks and commodities are partly related to the power law
probability distribution of returns and partly to the higher order
temporal correlations present.


\subsubsection*{Acknowledgments}

\noindent
We thank L.~A.~N.~Amaral, P.~Gopikrishnan, V.~Plerou, A.~Schweiger for
helpful discussions and suggestions and BP and NSF for financial
support. Y. A. thanks the Bikura fellowship for financial support.




\eject
\vspace*{0.5cm}

\eject

\vspace*{0.5cm}

\eject

\vspace*{0.5cm}

\begin{references}

\bibitem{Books} J.~P.~Bouchaud and M.~Potters, {\it Theory of Financial
Risk\/} (Cambridge University Press, Cambridge 2000); R.~N.~Mantegna and
H.~E.~Stanley, {\it An Introduction to Econophysics: Correlations and
Complexity in Finance\/} (Cambridge University Press, Cambridge 2000).

\bibitem{Works} M.~M. Dacorogna {\it et al.}, J. Int'l Money and Finance
{\bf 12}, 413 (1993); G.~Weisbuch {\it et al.}, Econ J.  {\bf 463}, 411
(2000); J.~P.~Nadal {\it et al.}, in {\it Advances in Self-Organization
and Evolutionary Economics}, edited by J. Lesourne and A. Orlian
(Economica, London, 1998), p. 149; P. Gopikrishnan, V. Plerou,
L. A. N. Amaral, M. Meyer, and H. E.  Stanley, Phys. Rev. E {\bf 60},
5305 (1999); Y. Liu, P. Gopikrishnan, P. Cizeau, M. Meyer, C.-K. Peng,
and H. E.  Stanley, Phys. Rev. E {\bf 60}, 1390 (1999); V. Plerou,
P. Gopikrishnan, L. A. N. Amaral, M. Meyer, and H. E.  Stanley,
Phys. Rev. E {\bf 60}, 6519 (1999);

\bibitem{Kaushik} K. Matia, L. A. N. Amaral, S. Goodwin, and
H. E. Stanley, Phys. Rev. E {\bf 66}, 045103 (2002).

\bibitem{Gene} H. E. Stanley and P. Meakin, Nature {\bf 335}, 405 (1988).

\bibitem{Barabasi} A.L. Barabasi and T. Vicsek, Phys. Rev. A {\bf 44},
2730 (1991).

\bibitem{Mandelbrot} B.~B.~Mandelbrot, SCI AM {\bf 280}, 70 (1999).

\bibitem{Bershadskii} A. Bershadskii, J. Phys. A: Math. Gen. {\bf 34},
L127 (2001).

\bibitem{Strutzik} Z. R. Struzik, Physica A {\bf 296}, 307 (2001).

\bibitem{Fisher} L.~Calvet and A.~Fisher, Rev. Econ. Stat. {\bf 84}, 381
(2002); {\it ibid.} J. Econometrics {\bf 105}, 27 (2001).

\bibitem{Sun} X. Sun, H. Chen, Z. Wu, and Y. Yuan, Physica A {\bf 291},
553 (2001).

\bibitem{Canessa} E. Canessa, J. Phys. A: Math. Gen. {\bf 33}, 3637 (2000).

\bibitem{Ausloos} M.~Ausloos and K.~Ivanova, Comput. Phys. Commun. {\bf
147}, 582 (2002); K.~Ivanova and M.~Ausloos, Eur. Phys. J. B {\bf 8},
665 (1999).

\bibitem{Bacry} E. Bacry, J. Delour, and J. F. Muzy, Phys. Rev. E {\bf
64}, 026103 (2001).

\bibitem{Pochart} B.~Pochart and J.-P.~Bouchaud, preprint cond-mat/0204047.

\bibitem{Jan} J.~W.~Kantelhardt, S. Zschiegner, E. Koscielny-Bunde,
A. Bunde, S. Havlin, and H. E. Stanley, preprint physics/0202070.

\bibitem{Cdatabase} See {\tt http://www.platts.com}.

\bibitem{Sdatabase} See {\tt http://finance.yahoo.com}.

\bibitem{Shuffling-note} We repeat the procedure of randomizing in 3
different ways: (i) Fourier phase randomization following
Schreiber~\cite{Schreiber}. In the phase randomization procedure phases
of a times series are randomize. This procedure thus preserves the two
point correlation but destroys any higher order correlation. (ii)
Randomizing the sign but preserving the ordering of the absolute
values. (iii) Randomizing the absolute values but preserving the
ordering of the signs of the time series. All the above procedure yields
similar results for the MF spectrum.

\bibitem{Peng} C. K. Peng, S. V. Buldyrev, S. Havlin, M. Simons,
H. E. Stanley and A.  L. Goldberger, Phys. Rev. E {\bf 49}, 1685 (1994);
the fluctuation function of DFA is $[F_{q=2}(s)]^{1/2}$.  See also
J. W. Kantelhardt, E. Koscielny-Bunde, H. H. A. Rego, S. Havlin, and
A. Bunde, Physica A {\bf 295}, 441 (2001); K. Hu, Z. Chen,
P. Ch. Ivanov, P. Carpena, and H. E. Stanley, Phys. Rev. E {\bf 64},
011114 (2001); Z. Chen, P. Ch. Ivanov, K. Hu, and H. E. Stanley,
Phys. Rev. E {\bf 65}, 041107 (2002).

\bibitem{Note2} To compare $\tau(q)$ for stocks and $\tau(q)$ for
commodities we use the Kolmogorov Smirnov (KS) test for each value of
$q$. For $\tau(q)|_{q<0}$ we find the KS probability, $P_{KS}$ is
$\approx 10^{-4}$, much less than 0.05, suggesting that $\tau(q)|_{q<0}$
for stocks is different from that for commodities. For $\tau(q)|_{q>0}$
$P_{KS}$ increases to $\approx 0.4$ (more than 0.05) suggesting that
stocks and commodities are statistically indistinguishable. We also
apply the receiver operative characteristic (ROC) \cite{Swets}, another
non-parametric analysis to test the degree of separation between
commodities and stocks. We find that the ROC results are consistent with
the KS results. The KS and ROC tests were repeated for each $q$ value on
the shuffled returns, and we find $\tau(q)$ for stocks and commodities
become statistically comparable since $P_{KS}\approx 0.6$.


\bibitem{Pfn-note} We also repeat the MF analysis in the following way:
(i) Evaluate $F_{q,i}(s)$ where $i=1,...29$ for commodities and
$i=1,...2449$ for stocks. (ii) Evaluate $F_{q,\mbox{\scriptsize av}}(s)
= \frac{1}{N}\sum_{i=1}^{N}F_{q,i}(s)$ where $N=29$ for commodities, and
$N=2449$ for stocks. (iii) Evaluate $\tau_{\mbox{\scriptsize av}}(q)$ from
$F_{q,\mbox{\scriptsize av}}(s) \sim s^{\tau_{\mbox{\scriptsize
av}}(q)}$. We observe that $\tau_{\mbox{\scriptsize av}}(q)$ and $f(h)$
evaluated in this procedure is similar to that obtained following the
method described in the text.

\bibitem{Swets} J.~A.~Swets, Science {\bf 240}, 1285 (1988).

\bibitem{Schreiber} T.~Schreiber and A.~Schmitz, Phys. Rev. Lett. {\bf
77}, 635 (1996); {\it ibid.} Physica D {\bf 142}, 346 (2000).


\end{references}
\end{document}